# Atomistic View of Homogeneous Nucleation of Water into Polymorphic Ices

Maodong Li[1,2], Jun Zhang[2], Niu Haiyang[3], Yao Kun Lei[2], Xu Han[4], Lijiang Yang[4], Zhiqiang Ye[1,2], Yi Isaac Yang[2*], and Yi Qin Gao[2,4*]

**Addresses**

[1]School of Chemical Biology and Biotechnology, Peking University Shenzhen Graduate School, Shenzhen 518055, China

[2]Institute of Systems Physical Biology, Shenzhen Bay Laboratory, Shenzhen 518132, China

[3]International Centre for Materials Discovery, School of Materials Science and Engineering, Northwestern Polytechnical University, Xi'an, Shaanxi 710072, China

[4]College of Chemistry and Molecular Engineering, Peking University, Beijing 100871, China

*Corresponding authors: Yi Isaac Yang (yangyi@szbl.ac.cn), Yi Qin Gao (gaoyq@pku.edu.cn)

**Abstract**

Water is one of the most abundant substances on Earth, and ice, i.e., solid water, has more than 18 known phases. Normally ice in nature exists only as Ice Ih, Ice Ic, or a stacking disordered mixture of both. Although many theoretical efforts have been devoted to understanding the thermodynamics of different ice phases at ambient temperature and pressure, there still remains many puzzles. We simulated the reversible transitions between water and different ice phases by performing full atom molecular dynamics simulations. Using the enhanced sampling method MetaITS with the two selected X-ray diffraction peak intensities as collective variables, the ternary phase diagrams of liquid water, ice Ih, ice Ic at multiple were obtained. We also present a simple physical model which successfully explains the thermodynamic stability of ice. Our results agree with experiments and leads to a deeper understanding of the ice nucleation mechanism.

Water-to-ice transition is still a mystery[1] which puzzles scientists working in biochemistry[2], physics[3] and environmental science[4-6]. Understanding the ice interaction with biomacromolecule is vital to functional protein design, such as antifreeze proteins[7], ice-nucleating proteins[8] or organic polymers[9]. The phase diagram of ice is complicated with more than eighteen solid phases[10], but only the hexagonal ice (ice Ih, ABAB... stacking[11]) and the cubic ice (ice Ic, ABCABC... stacking[11]) are present in nature. At ambient pressure, water freezes into Ih rather than Ic[12]. Ice Ic only exists partially at low temperatures and high pressure[13]. In fact, pure ice Ic without stacking defects was not experimentally accessible until 2020[10]. In most cases, water freezes into stacking disordered ice (ice Isd) consisting of hexagonal and cubic structures[14-18].

Molecular dynamics (MD) simulation is an efficient tool in seeking the ice stacking mechanism on the molecular level. However, with the limitation imposed by the large conformation space and the timescale[19,20] in crystallization, a direct all-atom simulation of ice nucleation from liquid water is almost impossible[21]. The only one successful case of spontaneous nucleation with all-atom model is the work of Ohmine and co-workers[22] in 2002, which has been difficult to be reproduced. Therefore, many groups used instead enhanced sampling methods to help simulate ice nucleation[23-28]. However, reversible transition between water and ice is still a great challenge. Recently, Niu and co-workers[29] realized the reversible water-ice transition in all-atom MD simulations using the enhanced sampling method MetaITS[30] with a carefully designed set of collective variables (CVs), which includes the X-ray diffraction (XRD) peak intensities[31]. The XRD peak intensity is an efficient CV for the phase transition between the solid and liquid states, and the simulation of reversible transition has been succeeded using the metadynamics with only one XRD peak intensity as CV for the systems of Na, Al[32], Si[33], etc. It turned out that the phase transition between water and ice is much more difficult than these covalent crystals. Although a complex combination of seven X-ray diffraction peak intensities and the translation entropy of water molecules were used as CVs, the solid state obtained in Ref. 28 is a mixed phase of stacking disordered ices. An efficient sampling of different ice phases using MD simulation thus remains as a challenge.

In this paper, we performed MD simulation with the all-atom TIP4P/Ice water model. The TIP4P/Ice model[34] provides one of the best predictions for the melting point in comparison to the experiment data, and was widely used in homogeneous nucleation of water[33,35,36]. Rather than the coarse-grained models, like monotonic water (MW) potential[23,25,37-41], the all-atom TIP4P/Ice model provides precise dynamics behaviours as well as H-bond connection and dipole properties. Our simulation system contains 1152 water molecules placed in a periodic box of 3.14 nm * 2.72 nm * 4.45 nm, which can match the lattice of ice Ih, Ic or their stacking disordered states (See Section S-I for detailed information).

We adopt the enhanced sampling methods MetaITS to facilitate the phase transition between water and ice. MetaITS[30] is a hybrid method that combines metadynamics[42] (MetaD) and integrated tempering sampling[43] (ITS), which increases the sampling efficiency of MD simulation by accelerating the CVs related degrees of freedom (See Section S-III for detailed information). The CVs we used in this work are the two intensities of 3D and 2D XRD peaks[33]:

$$s_1(\mathbf{R}) = s_{\text{XRD}}(Q(\theta_1); \mathbf{R}) = \frac{1}{N} \sum_{i=1}^{N} \sum_{j=1}^{N} f_i(Q) f_j(Q) \frac{\sin(Q \cdot R_{ij})}{Q \cdot R_{ij}} \omega(R_{ij}), \quad (1)$$

$$s_2(\mathbf{R}) = s_{\text{XRD}}^{xy}(Q(\theta_2); \mathbf{R}) = \frac{1}{N} \sum_{i=1}^{N} \sum_{j=1}^{N} f_i(Q) f_j(Q) J_0(Q \cdot R_{ij}^{xy}) \omega^{xy}(R_{ij}^{xy}) \omega^z(R_{ij}^z), \quad (2)$$

in which $Q = \frac{4\pi \sin\theta}{\lambda}$ is the scattering vector, where $\lambda$ = 0.15406 nm is the wavelength of the X-ray[29], and $\theta$ is the diffraction angle, in Eq.(1-2), $\theta_1$ = 5.98° and $\theta_2$ = 5.65°. $f_i(Q)$ and $f_j(Q)$ are the atomic scattering form factors and $R_{ij}$ is the distance between atoms $i$ and $j$. $\omega(R_{ij}) = \dfrac{\sin(\pi R_{ij}/R_c)}{\pi R_{ij}/R_c}$ is the window function. $R_{ij}^{xy}$ is the distance between atom $i$ and $j$ of the projection in the x-y plane, and $J_0(Q \cdot R_{ij}^{xy})$ is the Bessel function. More details are shown in Section S-II in SI.

We first performed the MD simulation at 270 K with MetaITS, which generates an effective potential that can be considered as a linear combination of a serial of Boltzmann distribution at a range of temperatures from 260 K to 280 K. See Section S-X in SI for detailed information. Six simulation trajectories totalling 28.8 μs were collected, and hundreds of transitions between water and ice are observed (See the Section S-V and video S1 in SI for detail information) which demonstrate the high efficiency of MetaITS with the two XRD-type CVs in sampling ice structures. Another advantage of MetaITS is that properties of the system in a large range of temperatures can be calculated at a cost comparable to a single MetaD run. Therefore, the free energies of water (ice percentage $I\% \leq 10\%$, defined in SI) $g_{\text{water}}$ and ice ($I\% \geq 90\%$) $g_{\text{ice}}$ and the free energy changes from water to ice $\Delta G_{\text{water} \to \text{ice}}$ are calculated and shown in Fig.1a. The melting point $T_m$ calculated from the intercept of the $\Delta G_{\text{ice} \to \text{water}}$ axis with the zero gives is 273.7 ± 0.7 K, in reasonable agreement with the reported melting point of the TIP4P-Ice model, 272.2 K[34]. Similarly, the difference in the entropy between ice and water $\Delta S_{\text{ice-water}}$ can also be calculated from the slope, which is -17.29 ± 0.5 J/(mol·K) and also fits well with the TIP4P/Ice model[34], -16.8 J/(mol·K). Some other thermodynamics properties of our simulation results are summarized in Section S-VI in SI.

Due to the anisotropy of the CV $s_2$, the plane of sphere packing can only stack at the Z-axis of the simulation system, and up to 12 ice bilayers can be formed at this size of periodic box. Since each bilayer may form a hexagonal or cubic structure, fully iced systems can form the states with different hexagonal/cubic ratios. In the simulation, seven different hexagonal/cubic ratios of structures were observed. The sampled structures as well as the corresponding cubicity (the fraction of cubic ice, using a ternary definition[44]) are shown Fig. 1(b-i) and Section S-VII in SI.

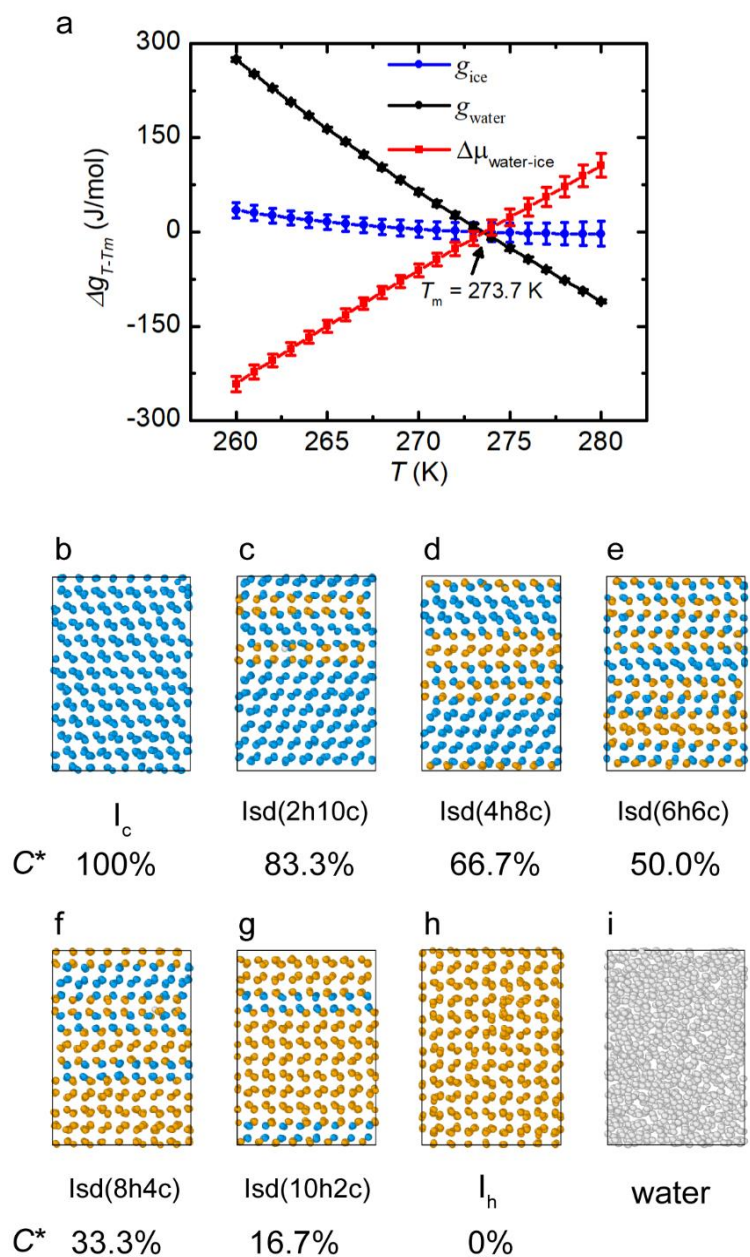

Figure 1 | (**a**) Temperature dependence of water-ice transitions. Free energy difference between water(black) and ice(blue) is shown as a function of temperature. The black line shows the linear dependence of $\Delta\mu_{water-ice}$ on the temperature at each reweighting temperature, of which the derivative represents the corresponding conformational

entropy. (**b** - **i**) Typical snapshots from the x-z plane of ice polymorphs and water, blue for Ic, orange for Ih and grey for liquid, more details given in Section S-VII. The compositions of ice polymorphs are also marked: Ic, Isd(2h10c, two hexagonal bilayers and ten cubic ones, the same below), Isd(4h8c), Isd(6h6c), Isd(8h4c), Isd(10h2c) and Ih, which corresponds to the fractions of cubic stacking sequences (cubicity, $C^*$).

The free energy surface (FES) as a function of the two CVs $s_1$ and $s_2$ is computed and shown in Fig. 2a. This figure shows that the CV $s_1$ can distinguish effectively the solid and liquid states, and the CV $s_2$ can distinguish well the different states of ice. Therefore, the usage of these two CVs allows efficient transitions between water and ice as well as an effective sampling of different ice states. There are 7 energy basins for the solid state on FES, which respectively correspond to the 7 ice structures with different values of cubicity (Fig. 1). We also calculated the ternary phase free energy diagram of the liquid state, ice Ih, and ice Ic (Fig.2b). Similar to the FES in Fig. 2a, this phase diagram also shows 7 ice states as distinguished by their different values of cubicity.

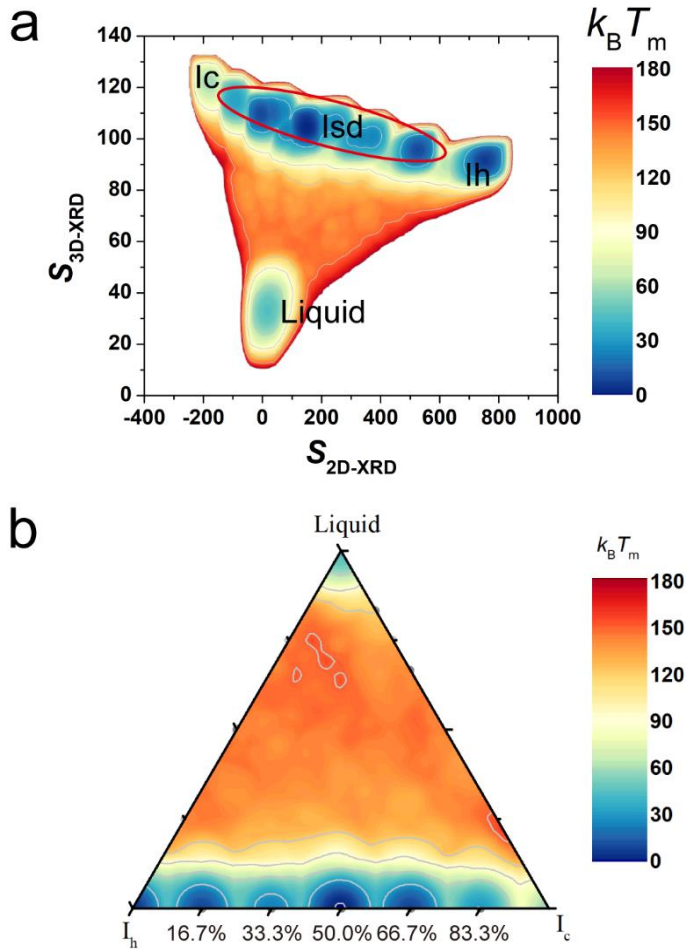

Figure 2 | Thermodynamics properties of simulated water-ice system. (**a**) Free energy surface in terms of collective variables at 270 K, $s_{\text{2D-XRD}}$ and $s_{\text{3D-XRD}}$. Each potential well is assigned to cluster identifiers as Fig.1. (**b**) The ternary phase free energy diagram at 270 K. The left axis corresponds to *H*% (Ih molecule percentage in the system, so as *W*% to water percentage and *C*% to Ic percentage, defined in Section S-IV in SI) which increases from top to bottom. The right axis corresponds to *W*% increasing from bottom to top. The bottom axis is corresponding to *C*% increases from left to right.

To further investigate the thermodynamics properties of the different ice state, we calculated the free energy change for water to ice transition, $\Delta G_{\text{water}\rightarrow\text{ice}}$, as a function of cubicity from 260K to 280K in Fig. 3a. This figure clearly exhibits that there are 7 stable states with the cubicity $C^* = 0\%$, 16.7%, 33.3%, 50%, 66.7%, 83.3% and 100%, which correspond to the seven states in Fig. 1: Ih, Isd(2h10c), Isd(4h8c), Isd(6h6c), Isd(8h4c), Isd(10h2c), and Ic, respectively.

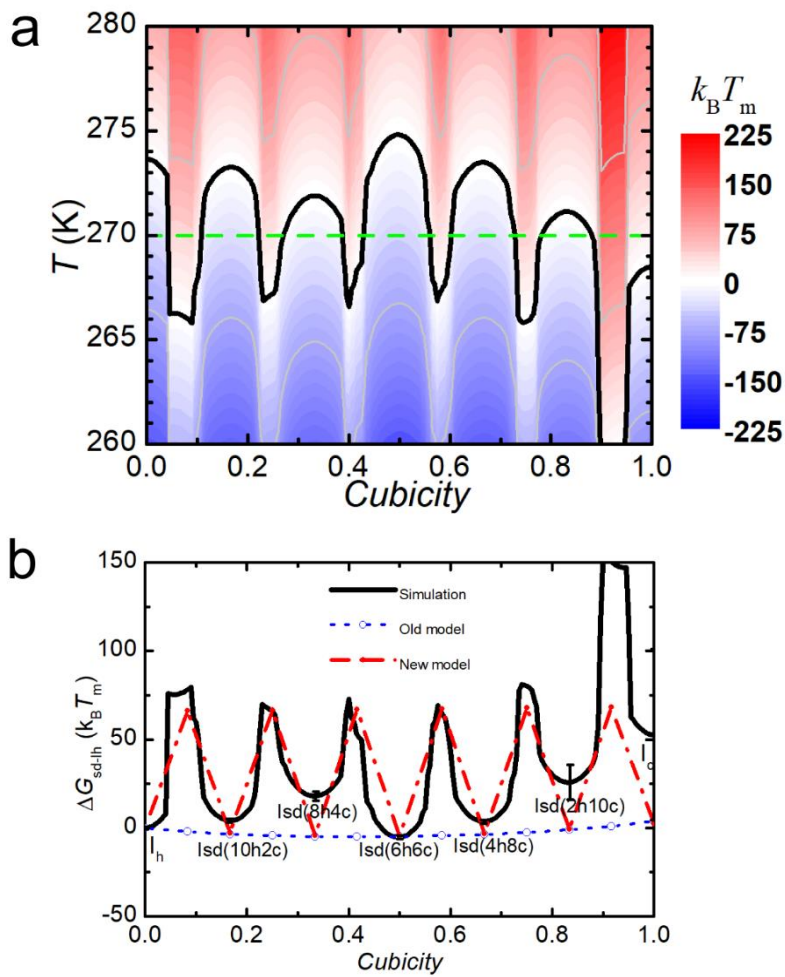

Figure 3 | Thermodynamical analysis of exquisite ice polymorphisms. (**a**) $\Delta G_{\text{water}\to\text{ice}}$ as a function of cubicity and temperature, using liquid phase in the corresponding temperature as the reference. The black line indicates the melting points of each polymorph. (**b**) Free energy plot as a function of cubicity, with Ih as the reference at 270 K. The stacking model is adopted to qualitatively describe the discrete distribution.

According to the theory of Lupi *et al*[23], whether any bilayer is hexagonal or cubic is independent of the neighbouring layers, so the system containing $N$ planes should have a total of $N+1$ possible combinations. However, only 7 different states can be found in our systems containing 12 planes. The current study revealed that the assumption above does not hold under the periodic boundary condition. Instead, a different scenario is suggested: whether an ice bilayer is hexagonal or cubic is determined by itself, but by the close packed structure of its two adjacent layers. In other words, we need to know the close packed structure of each bilayer before we can determine which bilayer is hexagonal or cubic: an ice bilayer is hexagonal if its two neighbouring ice bilayers are in the same close-packed structure (A-B-A stacking); conversely it is cubic if its two neighbouring ice bilayers are in different close-packed structures (A-B-A stacking). It is important to note that this model needs to ensure that the structures of two adjacent bilayers cannot be the same, so the number of reasonable arrangements of ice bilayers at the periodic boundary condition is much smaller than in the non-periodic system. For example, in the periodic system containing 12 ice bilayers, a total of 683 possible arrangements of the close-packed structure (A, B or C) can be formed, which corresponds to 7 combinations of hexagonal and cubic bilayers. See Section S-VIII in SI for detailed information.

According to the model discussed above, a simple estimate can be obtained for the free energy of different states of ice:

$$\Delta G_{\text{sd-Ih}} = \Delta H_{\text{cubicity}} + \Delta H_{\text{surface}} + \Delta H_{\text{fault}} - T\Delta S_{\text{layer}}. \quad (3)$$

where $\Delta H_{\text{cubicity}}$ is the free energy difference between bulk hexagonal and cubic ices[11], $\Delta H_{\text{surface}}$ is the free energy difference between ice and liquid water[45], which is zero pure solid ice, $H_{\text{fault}}$ is the penalty of the mismatched interface, $\Delta S_{\text{layer}} = R \times \ln(\Omega(C^*))$ by is the entropy of layer, and $\Omega(C^*)$ is the number of combinatory way of hexagonal and cubic layers in the system. The main difference between the current and earlier theories is in the means of $\Omega(C^*)$ calculation. It appears that the current model is in clearly better consistency with the simulation results. We calculated the free energy of ice as a function of cubicity at 270 K using Eq. 3 and show the results in Fig. 3b. Compared to the model of Lupi *et al.*, the current model is more consistent with the simulations: It accurately predicts that 7 stable states of ice can be formed in the simulation system, while the old model predicts 13 states. Furthermore, our model also correctly predicts that the ice Isd(6h6c) is the most stable state, followed by ice Ih, and Ice Ic is the most unstable state. This simple model, however, has difficulties to precisely describe the stability of ice states, such as ice Isd(8h4c), the explanation of which might involve a more detailed molecular model.

In this computational study, we investigated the various states of ice, including ice Ih, ice Ic and various states of stacking disordered ice, around melting temperature and pressure with TIP4P/ice all-atom model. These structures were sampled efficiently using the enhanced sampling method MetaITS with the help of two XRD-based CVs. A number of thermodynamic properties including the ternary phase diagram are calculated from the simulation trajectories, which agree with literature. Our simulations also show that in the simulation system containing a maximum of 12 ice bilayers, seven, but not thirteen (as predicted by earlier theories), stable ice states can exist. We propose an alternative model, which satisfactorily explains the stable structure of ice that may exist in the system, especially in the periodic boundary conditions. We also propose a method to estimate the free energy of various states of ice based on this simple model. The corresponding calculation provides a qualitatively explanation on the stability of different states of ice in the simulation system. Further studies, including those on the kinetics of ice nucleation and simulation of larger systems etc., are currently underway, which hopefully will lead to further understanding of the fascinating physical properties of ice.

**REFERENCES**

1   Bartels-Rausch, T. Ten things we need to know about ice and snow. *Nature* **494**, 27-29, doi:10.1038/494027a (2013).
2   Hudait, A., Odendahl, N., Qiu, Y., Paesani, F. & Molinero, V. Ice-Nucleating and Antifreeze Proteins Recognize Ice through a Diversity of Anchored Clathrate and Ice-like Motifs. *J Am Chem Soc* **140**, 4905-4912, doi:10.1021/jacs.8b01246 (2018).
3   Bartels-Rausch, T. *et al.* Ice structures, patterns, and processes: A view across the icefields. *Rev Mod Phys* **84**, 885-944, doi:10.1103/RevModPhys.84.885 (2012).
4   Herbert, R. J., Murray, B. J., Dobbie, S. J. & Koop, T. Sensitivity of liquid clouds to homogenous freezing parameterizations. *Geophys Res Lett* **42**, 1599-1605, doi:10.1002/2014gl062729 (2015).
5   Coluzza, I. *et al.* Perspectives on the Future of Ice Nucleation Research: Research Needs and Unanswered Questions Identified from Two International Workshops. *Atmosphere* **8**, 138 (2017).
6   Kiselev, A. *et al.* Active sites in heterogeneous ice nucleation—the example of K-rich feldspars. *Science* **355**, 367-371, doi:doi:10.1126/science.aai8034 (2017).
7   Hudait, A. *et al.* Preordering of water is not needed for ice recognition by hyperactive antifreeze proteins. *Proceedings of the National Academy of Sciences* **115**, 8266-8271, doi:10.1073/pnas.1806996115 (2018).
8   Qiu, Y., Hudait, A. & Molinero, V. How Size and Aggregation of Ice-Binding Proteins Control Their Ice Nucleation Efficiency. *Journal of the American Chemical Society* **141**, 7439-7452, doi:10.1021/jacs.9b01854 (2019).
9   Naullage, P. M. & Molinero, V. Slow Propagation of Ice Binding Limits the Ice-Recrystallization Inhibition Efficiency of PVA and Other Flexible Polymers. *Journal of the American Chemical Society* **142**, 4356-4366, doi:10.1021/jacs.9b12943 (2020).
10  del Rosso, L. *et al.* Cubic ice Ic without stacking defects obtained from ice XVII. *Nature Materials* **19**, 663-668, doi:10.1038/s41563-020-0606-y (2020).
11  Quigley, D. Communication: Thermodynamics of stacking disorder in ice nuclei. *The Journal of Chemical Physics* **141**, 121101, doi:10.1063/1.4896376 (2014).
12  Bielska, K. *et al.* High-accuracy measurements of the vapor pressure of ice referenced to the triple point. *Geophysical research letters* **40**, 6303-6307, doi:10.1002/2013gl058474 (2013).
13  Murray, B. J., Knopf, D. A. & Bertram, A. K. The formation of cubic ice under conditions relevant to Earth's atmosphere. *Nature* **434**, 202-205, doi:10.1038/nature03403 (2005).
14  Kuhs, W. F., Sippel, C., Falenty, A. & Hansen, T. C. Extent and relevance of

stacking disorder in "ice I(c)". *Proc Natl Acad Sci U S A* **109**, 21259-21264, doi:10.1073/pnas.1210331110 (2012).

15  Malkin, T. L., Murray, B. J., Brukhno, A. V., Anwar, J. & Salzmann, C. G. Structure of ice crystallized from supercooled water. *Proc Natl Acad Sci U S A* **109**, 1041-1045, doi:10.1073/pnas.1113059109 (2012).

16  Malkin, T. L. *et al.* Stacking disorder in ice I. *Phys Chem Chem Phys* **17**, 60-76, doi:10.1039/c4cp02893g (2015).

17  Amaya, A. J. *et al.* How Cubic Can Ice Be? *J Phys Chem Lett* **8**, 3216-3222, doi:10.1021/acs.jpclett.7b01142 (2017).

18  Morishige, K. & Uematsu, H. The proper structure of cubic ice confined in mesopores. *J Chem Phys* **122**, 044711, doi:10.1063/1.1836756 (2005).

19  Haji-Akbari, A. & Debenedetti, P. G. Direct calculation of ice homogeneous nucleation rate for a molecular model of water. *Proceedings of the National Academy of Sciences of the United States of America* **112**, 10582-10588, doi:10.1073/pnas.1509267112 (2015).

20  Mayer, E. & Hallbrucker, A. Cubic ice from liquid water. *Nature* **325**, 601, doi:10.1038/325601a0 (1987).

21  Vrbka, L. & Jungwirth, P. Homogeneous Freezing of Water Starts in the Subsurface. *The Journal of Physical Chemistry B* **110**, 18126-18129, doi:10.1021/jp064021c (2006).

22  Matsumoto, M., Saito, S. & Ohmine, I. Molecular dynamics simulation of the ice nucleation and growth process leading to water freezing. *Nature* **416**, 409-413, doi:10.1038/416409a (2002).

23  Lupi, L. *et al.* Role of stacking disorder in ice nucleation. *Nature* **551**, 218-222, doi:10.1038/nature24279 (2017).

24  Quigley, D. & Rodger, P. M. Metadynamics simulations of ice nucleation and growth. *J Chem Phys* **128**, 154518, doi:10.1063/1.2888999 (2008).

25  Reinhardt, A. & Doye, J. P. Free energy landscapes for homogeneous nucleation of ice for a monatomic water model. *J Chem Phys* **136**, 054501, doi:10.1063/1.3677192 (2012).

26  Reinhardt, A., Doye, J. P., Noya, E. G. & Vega, C. Local order parameters for use in driving homogeneous ice nucleation with all-atom models of water. *J Chem Phys* **137**, 194504, doi:10.1063/1.4766362 (2012).

27  Piaggi, P. M., Valsson, O. & Parrinello, M. Enhancing Entropy and Enthalpy Fluctuations to Drive Crystallization in Atomistic Simulations. *Phys Rev Lett* **119**, 015701, doi:10.1103/PhysRevLett.119.015701 (2017).

28  Radhakrishnan, R. & Trout, B. L. Nucleation of hexagonal ice (Ih) in liquid water. *J Am Chem Soc* **125**, 7743-7747, doi:10.1021/ja0211252 (2003).

29  Niu, H., Yang, Y. I. & Parrinello, M. Temperature Dependence of Homogeneous Nucleation in Ice. *Physical review letters* **122**, 245501, doi:10.1103/PhysRevLett.122.245501 (2019).

30  Yang, Y. I., Niu, H. & Parrinello, M. Combining Metadynamics and

   Integrated Tempering Sampling. *J Phys Chem Lett* **9**, 6426-6430, doi:10.1021/acs.jpclett.8b03005 (2018).

31 Murray, B. J. & Bertram, A. K. Formation and stability of cubic ice in water droplets. *Phys Chem Chem Phys* **8**, 186-192, doi:10.1039/B513480C (2006).

32 Zhang, Y. Y., Niu, H., Piccini, G., Mendels, D. & Parrinello, M. Improving collective variables: The case of crystallization. *The Journal of chemical physics* **150**, 094509, doi:10.1063/1.5081040 (2019).

33 Niu, H., Piaggi, P. M., Invernizzi, M. & Parrinello, M. Molecular dynamics simulations of liquid silica crystallization. *Proc Natl Acad Sci U S A* **115**, 5348-5352, doi:10.1073/pnas.1803919115 (2018).

34 Abascal, J. L., Sanz, E., García Fernández, R. & Vega, C. A potential model for the study of ices and amorphous water: TIP4P/Ice. *J Chem Phys* **122**, 234511, doi:10.1063/1.1931662 (2005).

35 Sanz, E. *et al.* Homogeneous ice nucleation at moderate supercooling from molecular simulation. *Journal of the American Chemical Society* **135**, 15008-15017, doi:10.1021/ja4028814 (2013).

36 Conde, M. M., Rovere, M. & Gallo, P. High precision determination of the melting points of water TIP4P/2005 and water TIP4P/Ice models by the direct coexistence technique. *The Journal of Chemical Physics* **147**, 244506, doi:10.1063/1.5008478 (2017).

37 Molinero, V. & Moore, E. B. Water modeled as an intermediate element between carbon and silicon. *J Phys Chem B* **113**, 4008-4016, doi:10.1021/jp805227c (2009).

38 Espinosa, J. R. *et al.* Interfacial Free Energy as the Key to the Pressure-Induced Deceleration of Ice Nucleation. *Phys Rev Lett* **117**, 135702, doi:10.1103/PhysRevLett.117.135702 (2016).

39 Espinosa, J. R., Vega, C., Valeriani, C. & Sanz, E. Seeding approach to crystal nucleation. *The Journal of Chemical Physics* **144**, 034501, doi:10.1063/1.4939641 (2016).

40 Cheng, B., Dellago, C. & Ceriotti, M. Theoretical prediction of the homogeneous ice nucleation rate: disentangling thermodynamics and kinetics. *Phys Chem Chem Phys* **20**, 28732-28740, doi:10.1039/c8cp04561e (2018).

41 Moore, E. B. & Molinero, V. Is it cubic? Ice crystallization from deeply supercooled water. *Phys Chem Chem Phys* **13**, 20008-20016, doi:10.1039/c1cp22022e (2011).

42 Laio, A. & Parrinello, M. Escaping free-energy minima. *Proceedings of the National Academy of Sciences of the United States of America* **99**, 12562-12566, doi:10.1073/pnas.202427399 (2002).

43 Gao, Y. Q. An integrate-over-temperature approach for enhanced sampling. *The Journal of Chemical Physics* **128**, 064105, doi:10.1063/1.2825614 (2008).

44  Maras, E., Trushin, O., Stukowski, A., Ala-Nissila, T. & Jónsson, H. Global transition path search for dislocation formation in Ge on Si(001). *Comput Phys Commun* **205**, 13-21, doi:https://doi.org/10.1016/j.cpc.2016.04.001 (2016).
45  Hudait, A., Qiu, S., Lupi, L. & Molinero, V. Free energy contributions and structural characterization of stacking disordered ices. *Phys Chem Chem Phys* **18**, 9544-9553, doi:10.1039/c6cp00915h (2016).

**Acknowledgements** The authors thank Yupeng Huang, Yijie Xia for useful discussion. Computational resources were supported by the Shenzhen Bay Lab Supercomputing Centre. This research was supported by the National Natural Science Foundation of China [22003042 to Y. I. Y.].

**Author Contributions** Y.Q.G., Y.I.Y. and J.Z. conceived the work. M.L. and Y.I.Y. performed and analysed all molecular simulations and calculations with the 1D stacking model. J.Z. provided codes for thermodynamic analysis on the ternary phase free energy diagram. N.H.Y. contributed the design of two CVs. M.L., J.Z., Y.K.L., X.H., L.Y., Z.Q.Y., Y.I.Y. and Y.Q.G. interpreted the results. Y.Q.G., Y.I.Y., J.Z. and M.L. wrote the paper.

**Author Information** Reprints and permissions information is available at www.nature.com/reprints. The authors declare no competing financial interests. Readers are welcome to comment on the online version of the paper. Publisher's note: Springer Nature remains neutral with regard to jurisdictional claims in published maps and institutional affiliations. Correspondence and requests for materials should be addressed to Y.Q.G. (gaoyq@pku.edu.cn) or Yi Isaac Yang (yangyi@szbl.ac.cn).